\documentclass[letterpaper,twocolumn,10pt]{article}
\usepackage{usenix,epsfig,endnotes}
\usepackage{booktabs} 
\usepackage{amssymb}
\usepackage{amsmath,amsthm}
\usepackage{graphicx}
\usepackage[tight,footnotesize]{subfigure}
\usepackage{multirow}
\usepackage{listings}
\usepackage{color}
\usepackage[colorlinks, linkcolor=red, anchorcolor=blue, citecolor=green]{hyperref}
\usepackage{hyperref}
\usepackage{caption}
\usepackage{algorithm}
\usepackage{algorithmicx}
\usepackage[noend]{algpseudocode}
\usepackage{xspace}
\usepackage{url}
\usepackage{breakurl}

\usepackage{flushend}

\newcommand{\pname}{Gosig\xspace}

\newcommand{\para}[1]{\smallskip\noindent\textbf{#1}}

\newcommand{\prm}{$PR$ message\xspace}
\newcommand{\ppm}{$P$ message\xspace}
\newcommand{\tcm}{$TC$ message\xspace}

\newcommand{\prms}{$PR$ messages\xspace}

\newcommand{\tcms}{$TC$ messages\xspace}

\newcommand{\mbs}{\texttt{max\_block\_size}\xspace}

\newcommand{\state}[4]{\langle #1,\allowbreak #2,\allowbreak #3,\allowbreak #4 \rangle\xspace}

\newtheorem{myLem}{Lemma}

\graphicspath{{figures/}}

\begin{document}

\title{\pname: Scalable Byzantine Consensus on Adversarial Wide Area Network for Blockchains}

\author{
{\rm Peilun Li}\\
Tsinghua University
\and
{\rm Guosai Wang}\\
Tsinghua University
\and
{\rm Xiaoqi Chen}\\
Princeton University
\and
{\rm Wei Xu}\\
Tsinghua University
} 





\maketitle

\begin{abstract}

Existing Byzantine fault tolerance (BFT) protocols face significant challenges in the consortium blockchain scenario.  On the one hand, we can make little assumptions about the reliability and security of the underlying Internet.  On the other hand, the applications on consortium blockchains demand a system as scalable as the Bitcoin but providing much higher performance, as well as provable safety.  
We present a new BFT protocol, \pname, that combines crypto-based secret leader selection and multi-round voting in the protocol layer with implementation layer optimizations such as gossip-based message propagation.  In particular, \pname guarantees safety even in a network fully controlled by adversaries, while providing provable liveness with easy-to-achieve network connectivity assumption.   
On a wide area testbed consisting of 140 Amazon EC2 servers spanning 14 cities on five continents, we show that \pname can achieve over 4,000 transactions per second with less than 1 minute transaction confirmation time.

\end{abstract}

%
%


\section{Introduction}
\label{intro}
The rise of cryptocurrencies, such as Bitcoin~\cite{nakamoto2008bitcoin}, increases the awareness and adoption of the underlying \emph{blockchain} technology.  Blockchain offers a distributed \emph{ledger} that serializes and records the transactions.  
Blockchain provides attractive properties such as full decentralization, offline-verifiability, and most importantly, scalable Byzantine fault tolerance on the Internet.  Thus, Blockchain has become popular beyond cryptocurrencies, expanding into different areas such as payment services, logistics, healthcare, and Internet-of-Things (IoT)~\cite{bonneau2015research,zheng2016blockchain,tschorsch2016bitcoin}.

While Bitcoin provides a \emph{permissionless} protocol, where everyone can join, we focus on \emph{consortium blockchains} (aka \emph{permissioned} blockchains), where a participant needs offline authentication to join.  It is useful in many commercial applications~\cite{guo2016blockchain}.  
While we no longer have to worry about Sybil attacks~\cite{douceur2002sybil}, there are still some other significant challenges.  

The blockchain is replicated to each participant, and the key problem is how to reach consensus on these replicas.  Comparing to traditional distributed transaction systems, the three biggest challenges of a blockchain are:

1) Players are from different organizations without mutual trust.  Failures, even Byzantine failures, are common.  Thus, we cannot rely on a small quorum (e.g.,  Chubby~\cite{burrows2006chubby} or ZooKeeper~\cite{hunt2010zookeeper}) for consensus. Instead, we need Byzantine consensus and allow all players to participate, i.e., supporting a consensus group with thousands of servers.

2) The system runs on an open Internet. 
The network can drop/delay communication arbitrarily.  Even worse, as the addresses of the participants are known to all, an adversary can easily launch attacks targeting any chosen participant at any time, using techniques like distributed denial-of-service (DDoS)~\cite{bitcoinDDoS,vasek2014empirical}.  The adversary can strategically choose which players to attack, and victims will remain unavailable until the adversary \emph{adaptively} choose to attack others. We call these attacks \emph{adaptive attacks}, which strongly threatens the special nodes in a protocol, such as the leaders.

3) Different from Bitcoin that allows temporary inconsistency (i.e., a \emph{fork}), people usually expect a permissioned chain to provide more traditional transaction semantics, i.e., a committed transaction is durable.  Also, applications on a permissioned chain~\cite{hyperledger} require much higher throughput and lower latency than Bitcoin.  

While there are many Byzantine fault-tolerant (BFT) protocols, most of them do not sufficiently address these three challenges.
For example, PBFT~\cite{castro1999practical}
and its successors~\cite{kotla2007zyzzyva, liu2015xft} and even some Bitcoin variants~\cite{eyal2016bitcoin} all depend on a leader or a small quorum to participate multiple rounds of communications, and thus they are vulnerable to \emph{adaptive attacks} on the leader. 
Other protocols that try to avoid faulty leaders by changing leaders in a fixed order~\cite{kwon2014tendermint, milosevic2013bounded,veronese2009spin,aiyer2005bar} cannot avoid adaptive leader attacks, because once the adversaries know who is the next leader, they can change their target accordingly.

There is a new generation of BFT protocols designed to run over the Internet.  
To avoid adaptive attacks,  Algorand~\cite{gilad2017algorand} hides the leader identity. 
To improve scalability, ByzCoin\cite{kogias2016enhancing} combines Proof of Work (PoW) with multi-signature-based PBFT.
To tolerate arbitrary network failures,
HoneyBadgerBFT\cite{miller2016honey} adopts asynchronous atomic broadcast~\cite{cachin2001secure} and asynchronous common subset (ACS)~\cite{ben1994asynchronous}.

Unfortunately, as we will detail in Section~\ref{keyfeatures}, none of these BFT protocols offer the following properties at the same time: 1) liveness under \emph{adaptive attack}, 2) scalability to 10,000s of nodes with low latency (15 seconds in our simulation) for commitment, and 3) provable safety (i.e., no fork at any time) with arbitrary network failure.  Also, there is no straightforward way to combine the techniques used in these protocols. 

We present \pname\footnote{The name is a combination of Gossip and Signature aggregation, two techniques we use.}, a new BFT protocol for permissioned blockchains.  \pname can achieve all three properties above, and also provide provable liveness with partially synchronous network (details in Section~\ref{systemmodel}).

\pname elects different leaders secretly for every block, and it eliminates the leader's involvement after it proposes a block to defend against adaptive attacks on leaders.
At the implementation level, we use gossip-based communications to fully exploit the link redundancy on the Internet while keeping the source safe.
Since we need to gather signatures during gossip, we adopt asynchronous multi-signature~\cite{boneh2003aggregate,syta2016keeping} to largely reduce the network overhead of consensus messages.

We evaluate \pname on an Amazon EC2-based 140-server testbed that spans 14 cities on five continents.  
We can achieve a throughput of 4,000 tps (transactions per second) with an average transaction confirmation latency of 1 minute.  
Even when 1/4 of the nodes fail, we can still maintain over 3,500 tps with the same latency.  
With over 100 participants, it effectively doubles the throughput and reduces the latency by 80\% comparing to HoneyBadgerBFT~\cite{miller2016honey}, the state-of-the-art protocol that offers the same level of safety guarantee.  
Also, using simulations, we show that \pname is able to scale to 10K nodes.  

In summary, our major contributions are: 

1) We propose a new BFT protocol that achieves scalability, provable safety and resilient to adaptive attack.  

2) We propose a novel method of combining \emph{secret leader selection, random gossip, multi-round voting}, and \emph{multi-signature} into a single BFT protocol in a compatible way.

3) We provide a real \pname implementation, evaluate it on a real-world geographically distributed testbed with 140 nodes, and achieve promising performance results.\footnote{We will opensouce \pname when the paper is published.}

\section{Related Work}

\para{Bitcoin and its variants.}
Permissionless public blockchains like Bitcoin~\cite{nakamoto2008bitcoin}, Ethereum~\cite{wood2014ethereum}, PPCoin~\cite{king2012ppcoin} need proof of work (PoW) or proof of stake (PoS) to prevent Sybil attacks. They also need incentive mechanisms to encourage people to join the public network to keep the system safe. 
Other designs~\cite{croman2016scaling,decker2016bitcoin} try to avoid chain forking but retain the design of PoW or PoS. We assume consortium blockchains~\cite{cachin2016architecture,swanson2015consensus,vukolic2015quest}, and mainly focus on the performance and safety of the system, instead of the other economic aspects. 

\para{Byzantine fault tolerance. }
The most important feature of a BFT protocol is safety.  Unfortunately, many open source BFT protocols are not safe~\cite{cachin2017blockchains}.  There are two major approaches to design provable BFT agreement protocols.  1) Using multi-round voting:  example systems include PBFT~\cite{castro1999practical} and its successors~\cite{kotla2007zyzzyva, liu2015xft,clement2009upright}; 2) Using leader-less atomic broadcast: HoneyBadgerBFT~\cite{miller2016honey} and~\cite{cristian1986atomic,kursawe2005optimistic}. 
To prevent malicious leaders from affecting the system, Aardvark~\cite{clement2009making} use performance metrics to trigger view changes  and Spinning~\cite{veronese2009spin}, Tendermint~\cite{kwon2014tendermint} or others~\cite{aiyer2005bar,milosevic2013bounded} rotates leader roles in a round robin manner. However, there methods can not avoid adaptive attacks because the leader role is known to all in advance, and thus can be muted by attacks like DDoS right before it becomes a leader.
\pname adopts similar voting mechanism like PBFT to get good performance without failure, and keeps safety and liveness under attacks.

In order to scale the system, many systems adopt the ``hybrid consensus'' design ~\cite{kiayias2017ouroboros,kogias2016enhancing,pass2017hybrid} that uses a Bitcoin-like protocol to select a small quorum, and use another (hopefully faster) BFT protocol to commit transactions.  If adversaries can instantly launch \emph{adaptive attacks} on leaders, it is hard for these protocols to maintain liveness.  Algorand~\cite{gilad2017algorand} leverages secret leader election and quorum member replacement methods to keep liveness.  \pname lets every player participate in the consensus, but combines similar secret leader selection with signature-based voting to prevent such attacks.




We use similar methods and adversary models proposed in Algorand~\cite{gilad2017algorand}. We adopt the idea of multi-round voting from PBFT and HoneyBadgerBFT~\cite{miller2016honey}, and the idea of multi-signature from ByzCoin~\cite{kogias2016enhancing}.  \pname combines these incompatible methods in a coherent protocol and achieves good performance.  We compare the key differences of these protocols in Section~\ref{keyfeatures}.

\para{Overlay network and gossip.}
Most BFT protocols and blockchains use broadcast as a communication primitive.  To improve broadcast reliability on the Internet, people often use application-layer overlay networks.  We adopt techniques like gossip from reliable multicast~\cite{birman1999bimodal}, probabilistic broadcast~\cite{eugster2003lightweight,kermarrec2003probabilistic} and other peer-to-peer (P2P) networks~\cite{karp2000randomized,venkataraman2006chunkyspread}.  Existing P2P networks may tolerate some Byzantine failures, but do not provide convergence guarantee~\cite{li2006bargossip}. By combining network optimizations like gossip with a robust protocol layer design, we can greatly improve both system resilience and availability.

\section{Problem Definition and Assumptions}
\label{problem_definition}


The goal of \pname is to maintain a blockchain.  
In \pname, clients submit \emph{transactions} to \emph{players} (or servers), who pack these transactions into \emph{blocks} in a specific order.
All committed blocks are then serialized as a \emph{blockchain}, which is is replicated to all players.  On the blockchain, one block extends another by including a hash of the previous block.
In a blockchain, a transaction is \emph{confirmed} only when the consensus group \emph{commits} the block containing the transaction. 
\pname, as a consensus protocol, ensures that all blockchain replicas are the same.  In particular, we want to prevent \emph{forks} on the blockchain, i.e. two different blocks extending the same block. 

\subsection{Problem Definition}
\label{sub:problem_definition}

We consider a system with $N$ players, $p_1, p_2, \cdots, p_n$.
We can tolerate $b$ static Byzantine failures and $c$ \emph{adaptive attacks} where $b+c= f = \lfloor (N-1)/3\rfloor$.  The Byzantine faulty nodes can do anything including colluding to break the protocol.  The honest players under \emph{adaptive attacks} act like crash failure, 
but they come to life again when the attacks are over. 
All other (at least $2f+1$) players are \emph{honest} and follow the protocol.  

All players form a \emph{consensus group} $\mathbf{G}$.
Each player $p_i$ outputs (commits) a series of ordered blocks 
$B_i[1], B_i[2],\allowbreak \cdots,\allowbreak B_i[n_i]$, where $n_i$, the length of the blockchain after attaching a new block, is the \emph{height} of this block. 

A transaction is an operation on a state machine, and we say it is valid when it is a legal operation on the current state.
A block $B_i[h]$ is valid if: 1) all transaction included are valid if executed sequentially,
and 2) the block header contains the correct reference to the previous block $B_i[h-1]$, like the Nakamoto  blockchain~\cite{nakamoto2008bitcoin}.

The goal of \pname is to let $\mathbf G$ reach a \emph{consensus on the blockchain}, i.e.
 the following two conditions hold.

%
%
%



\textbf{1. Safety}: 
(1) Any block committed by any honest player is valid;
(2) at any time, for any two honest players $p_i$ and $p_j$,
$B_i[k] = B_j[k]$ for any $k \leq \min(n_i, n_j)$.

\textbf{2. Liveness}:
For any finite time $t$ and any honest player $p_i$, 
there exists a time $t' >t$ when $p_i$ commits a block packed by an honest player.

Here we define safety and liveness of blocks instead of transactions for simplicity. We rely on gossip mechanisms to ensure that a transaction will reach most players, and an honest player will pack the transaction when it becomes the leader. Intuitively, the safety condition requires a total order of committed blocks on the blockchains
at all honest players, meaning there is no fork at any time.
The liveness condition says that all honest players will always make progress, i.e., if a transaction can reach all honest players, it will eventually be confirmed.  Both conditions are based on certain assumptions about the system, and we detail them next. 

\subsection{System Model and Assumptions}
\label{systemmodel}

We summarize our key assumptions that we use to prove the safety and liveness of \pname.  

\para{Strong cryptography and PKI.  } We only consider permissioned chain, and there is a trusted public key infrastructure (PKI) to authenticate each player - a common assumption in today's Internet.  We also assume the correctness of the cryptographic primitives. These assumptions are the foundation of the safety in \pname.


%


\para{Asynchronous network for safety.} 
Our protocol can keep safety under \emph{asynchronous network}~\cite{miller2016honey} condition, which means messages can be arbitrarily dropped, reordered, or duplicated.

\para{Liveness under partial synchrony.}
We also guarantee \emph{liveness} if the network has \emph{partial synchrony}, a common assumption~\cite{clement2009upright,kotla2007zyzzyva,milosevic2013bounded,veronese2009spin}.
We say a network has partial synchrony if
there exists a time $t'$ such that for any time $t>t'$, all messages sent between any two honest players during the interval can be delivered within a limited time bound $\Delta_t$. 
Specially, we assume \emph{adaptive attacks} on any player only take effect after a delay of $\Delta_t$ at any time. 


\para{Partially synchronized clock.} Similar to~\cite{gilad2017algorand}, we assume a partially synchronized clock for getting liveness.
That is, at any wall clock time $t$, for any two players $p_i$ and $p_j$, their local
time readings $C_{p_i}(t)$ and $C_{p_j}(t)$ satisfy that $| C_{p_i}(t) - C_{p_j}(t) | < \Delta$. 
Practically, it is easy to 
ensure a $\Delta$ of several seconds using standard Network Time Protocol (NTP) on the Internet.

\subsection{Key Features of \pname }
\label{keyfeatures}

Comparing to existing blockchains and other Byzantine-fault-tolerant atomic ordered broadcast protocols, \pname has achieved \emph{scalability}, \emph{liveness under adaptive attacks} and \emph{safety under asynchronous networks} at the same time.  While existing protocols provide one or more of these features, to our knowledge, \pname is the first protocol that offers all three together. 


ByzCoin~\cite{kogias2016enhancing} offers excellent scalability by combining PBFT, signature collection and proof-of-work (PoW).  However, like PBFT, it loses liveness under adaptive attacks given that it is still PBFT-based. 
Even without PBFT, its two-phase multi-signature and Bitcoin-NG~\cite{eyal2016bitcoin}-like mechanism that allows elected leaders to keep serving are also vulnerable to adaptive attacks. 

Algorand~\cite{gilad2017algorand} has excellent scalability, and tolerate adaptive attack using secret consortium election.  However, the safety of its Byzantine Agreement is based on a \emph{weak synchrony} assumption about the network.
This additional requirement comes from the idea of randomly selecting a small quorum, which is the key to Algorand's scalability.
We only adopt the secret \emph{leader} election from Algorand to avoid adaptive attacks, but completely redesign the BFT protocol using multi-round signature collections to achieve provable safety in asynchronous networks, like PBFT. We solve the scalability problem by combing protocol design with implementation optimizations like multi-signatures.

HoneyBadgerBFT~\cite{miller2016honey} achieves provable optimal liveness and provable safety in any situation. However, each node needs to send $O(N^2)$ messages per round.  Batching up $O(N^2\log N)$ transactions per round helps amortize the cost, but the large batch results in a latency as high as $O(N^2\log N)$, limiting the scalability.  In comparison, the network overhead for each \pname player is $O(N\log N)$ per round.  
Therefore, experimentally, we can double the HoneyBadgerBFT throughput with only 1/5 of the latency on a similar testbed with more participates.



In summary, we insist that \pname has provable safety under a strong adversary model, but we choose to relax the liveness goal a little, in exchange for better scalability.  We achieve this goal by adopting some originally incompatible ideas and provide alternative implementations, so they can be combined seamlessly with our accordingly designed protocol.

\section{\pname Protocol Overview}
\label{protocol}

We provide an intuitive overview of \pname and leave formal descriptions and analysis to Section~\ref{protocol_details}.

\para{Players.}
Every players participates in the protocol, and knows all other players' public key. It receives transactions submitted by clients and gossip transactions among all players. An honest player is responsible for verifying transaction validity and block validity. A player drops invalid transactions and blocks, and blacklist the senders. 
%
%
%
%
%

\begin{figure}[tb]
	\begin{center}
		\includegraphics[width=0.43\textwidth]{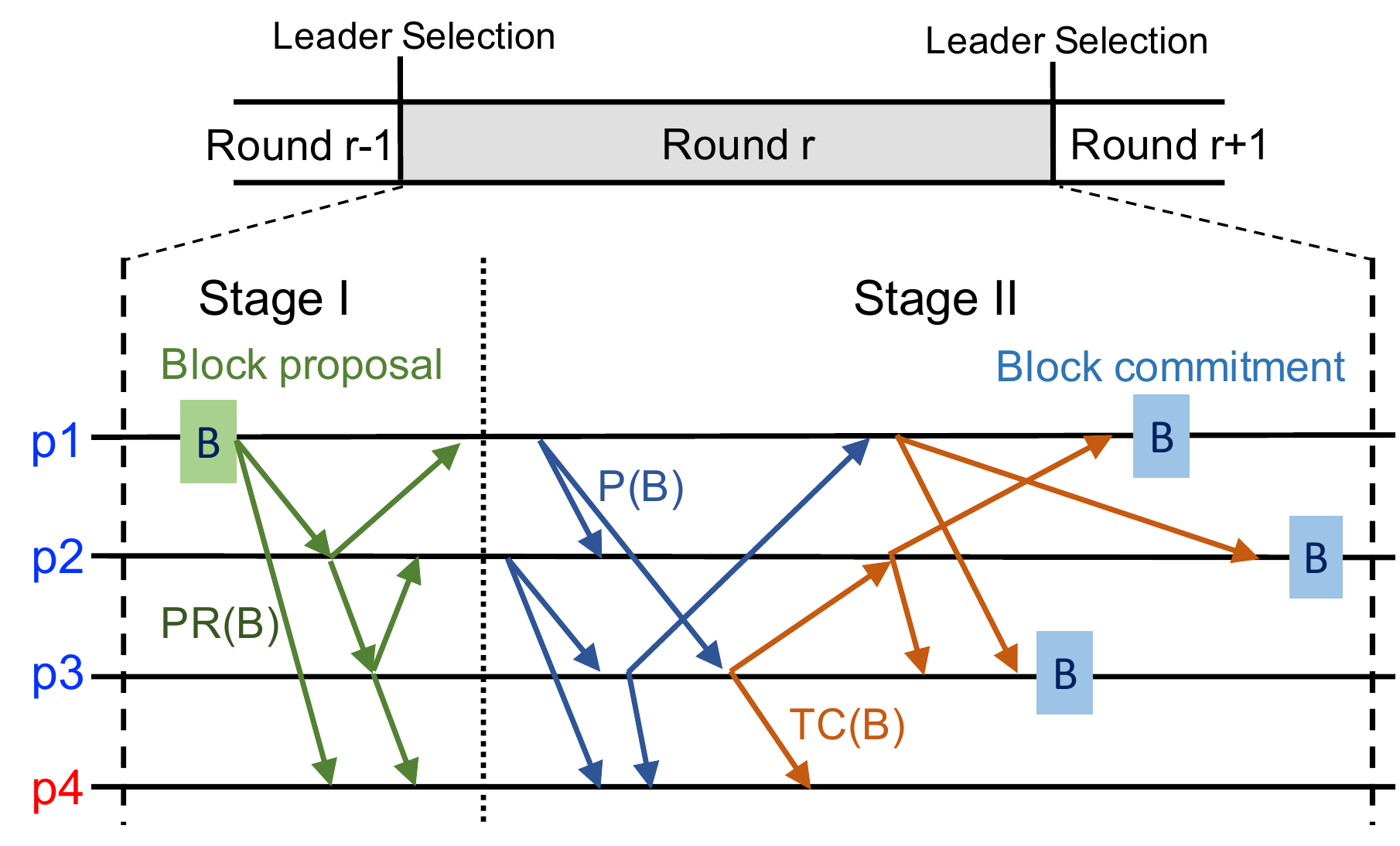}
		\caption{
			Overview of a \pname round (happy path only).}
		\label{fig:timeline}
	\end{center}
\end{figure}

\para{Rounds and stages.}
\pname is a partially-synchronous protocol.
We divide the execution of \pname into \emph{rounds} with a fixed time length (30 seconds by default).
Each round consists of a leader selection step (no communications) and two subsequent \emph{stages} with fixed length. 
Thus, all players know the current round number and stage by referring to the local clock.

Figure~\ref{fig:timeline} provides an overview of a typical round.  At the start of each round, some players secretly realize they are \emph{potential leaders} of this round with the cryptographic sortition algorithm (Section~\ref{leader_election}). Thus, the adversary can not target the leaders except randomly guessing.  

At Stage I, a selected potential leader packs non-conflict uncommitted transactions into a block proposal,
disseminates it with gossip, and acts just like a normal node afterwards.
Note that a potential leader's identity can only be discovered by others (including the adversary)
after she has full-filled her leader duties.


The goal of Stage II is to reach an agreement on some block proposal of this round by vote exchange.
A player ``votes'' for a block by adding her digital signature of the block to a \textsf{prepare} message  (``\ppm'') and sending it.  An honest player only votes for a single proposal per round. 
Upon receiving at least $2f + 1$ signatures from \ppm{}s for a block, the player starts sending \textsf{tentatively-commit} messages (``\tcm'') for it.  
She finally commits the block to her local blockchain replica once she receives $2f + 1$ \tcms.  

The above process only covers the ``happy path'' of the protocol. 
We provide the details how \pname handles failures in the next section. 

\section{\pname Protocol Details}
\label{protocol_details}

We formally describe the \pname protocol details.  
Throughout the paper, we adopt the following notation: subscripts denote who have signed the message\footnote{A message may be collectively signed by multiple players.} and superscripts denote the round in which the message is sent.  
For example, 
$P_X^r$ is a $P$ message collectively signed by the set $X$ of players in round $r$\footnote{For brevity, we denote $M_{\{i\}}^r$ by $M_i^r$.}.
We also use a shorthand to describe the action that $p_i$ sends a \ppm about a block $B$ as ``$p_i$ $P$s $B$'', and the action that $p_i$ sends a \tcm about a block $B$ as ``$p_i$ $TC$s $B$''.

\subsection{Player's Local State}
\label{local_state}

We describe each player as a finite state machine in \pname.  Each player maintains a local state
and decides her next actions based on the state and external events (receiving a certain message or selected as a leader).

The local state on player $p_i$ is a 4-tuple 
$s_i = \state{B_{root}}{h_{root}}{B_{tc}}{F}$.
The tuple contains two types of information.  The first two values are about the last committed block in her local blockchain copy.  $B_{root}$ is the block itself, and $h_{root}$ is the \emph{height} of $B_{root}$.  

The rest values in $s_i$ describe the \emph{pending block} that the player plans to add to her chain at height $h_{root} + 1$.  $B_{tc}$ is the pending block itself.  $B_{tc}$ is non-null if $p_i$ has $TC$ed $B_{tc}$.  Otherwise, $B_{tc}$ is a null value $\epsilon$.  
The last variable $F$ characterizes the timeliness of the block of $B_{tc}$.


Besides the elements appearing in the tuple, $p_i$ also implicitly maintains two variables:
$\overline {c_{root}}$ and $c_{tc}$.
$\overline {c_{root}}$ is a \emph{commitment certificate} that proves the validity of $B_{root}$, and
$c_{tc}$ is a \emph{proposal certificate} of $B_{tc}$.  We define both later in this section.
For brevity, we do not explicitly include $\overline {c_{root}}$ and $c_{tc}$ in the $s_i$ tuple.

When \pname starts running, a player's local state is initialized as 
$s_i = \langle B_{\epsilon}, 0, \epsilon, 0\rangle$.



\subsection{Leader Selection: The First Step}
\label{leader_election}

Leader selection is the first step of each round.
The objective of this step is to secretly and randomly select the \emph{potential leaders}
who are entitled to propose a next block.
The greatest challenge is to keep the election result unpredictable
until the potential leaders have sent out the \textsf{propose} messages (\prm). 
Otherwise, the adversary can attack the leaders beforehand, 
and thus break the liveness of the system.

\para{The cryptographic sortition algorithm.}
We use a simplified version of the \emph{cryptographic sortition} mechanism from Algorand~\cite{gilad2017algorand} to select a set of \emph{potential leaders}. 
Similarly, we use a number $Q^h$ to implement cryptographic sortition.
In \pname, $Q^h$ is recursively defined as
\begin{equation}
	Q^h = H(SIG_{l^h}(Q^{h-1})) \quad (h > 0)\\
\end{equation}
where $h$ is the height of a committed block $B$\footnote{
$Q^0$ is a random number shared among all players. 
}, 
$H$ is a secure hash function shared by all players,
$l^h$ is defined as the leader who has proposed the block $B$ 
(the signer of the proposal certificate of $B$),
and $SIG_i(M)$ is the digital signature of message $M$ signed with $p_i$'s private key.

Based on $Q^h$, we define a player $p_i$'s \emph{leader score} $L^r(i)$ at round $r$ as
$L^r(i) = H(SIG_i(r, Q^{h}))$,
where $h$ is the height of the latest committed block at round $r$.

At the beginning of each round $r$, each player $p_i$ computes her $L^r(i)$, and if 
the score is less than a \emph{leader probability parameter} $q$, she knows that she is a 
\emph{potential leader} of the round.  A potential leader can prove to other players about her leader status with the value $SIG_i(r, Q^h)$.  We define the value as the \emph{leader proof} for round $r$, $lc_i^r$.  
The process requires no communication among players. 


The cryptographic sortition algorithm has two good properties: 
1) the signature $SIG_i$ uses $p_i$'s private key, and thus cannot be forged by others; 
2) If the hash function $H(\cdot)$ is perfectly random, the potential leader selection is uniformly random.  Thus, there is no way for the adversary to know who is selected, nor can it change any node's chance of becoming a leader.

Choosing a right $q$ is important.  If $q$ is too small, some rounds may not have a leader and thus fail to proceed.  
If $q$ is too large, there may be many competing potential leaders, wasting resources to resolve the conflict.  Similar to Algorand, we set $q = 7/N$ where $N$ is the total number of players. This $q$ is sufficiently large to reduce the probability of no-leader rounds to less than 0.1\%.

Of course, a faulty node may still become a potential leader.  In this case, the worst damage it can cause is to stall the round without affecting the correctness.  

\subsection{Potential Leaders Propose Blocks}

When a player $p_i$ recognizes that she is a potential leader, she needs to decide which block
to propose and then generate a proposal message. 

Given $p_i$'s current local state $s_i = \state{B_{root}}{h_{root}}{B_{tc}}{F}$ ($B_{tc}$ may be $\epsilon$), she first gathers a list of candidate blocks to propose.  
If $p_i$ finds she has a non-empty $B_{tc}$, the block $B_{tc}$ is a good candidate to propose again, and the height stays the same as $B_{tc}$.
Alternatively, she can also construct a new block extending her own chain, at height ($h_{root} + 1$).  The following procedure defines which block she will propose.

To make a proposal valid, a potential leader needs to provide a valid \emph{certificate} $c$ for the proposed block $B$ at height $h$.
In addition to serving as a proof of the block validity, $c$ also determines the \emph{proposal round} $r_p(B)$\footnote{More precisely, the proposal round is an attribute of a proposal,
which is defined in the next paragraph, rather than of a block. 
We use $r_p(B)$ notation for brevity.} of the block proposal.  
The leader decides which candidate block to propose based on $r_p(B)$.

We can understand the proposal round as the round number when the block is first generated.  
Intuitively, a block with a larger proposal round is more likely to get accepted by peers. 
Therefore, in \pname, we stipulate that a potential leader should propose the block with 
\emph{the largest proposal round} among all candidate blocks.



A proposal certificate $c$ is \emph{valid} if and only if it matches one of the following two cases.  We also specify how we compute the proposal round in each case.

\noindent\textbf{Case 1}: 
$c = TC_i^{r'}(B^{r'}, h)$ $(r' < r)$, 
where $B$ and $h$ are exactly the proposed block and its height, respectively.
In this case, $r_p(B) = r'$;
	
	
\noindent\textbf{Case 2}: 
$c = TC_X^{r'}(B'^{r'}, h-1)$ $(r' < r)$, where $X$ contains at least $2f+1$ different players.
In this case, $r_p(B) = r' + 1$.

%
%
%


%
%
%
%

Finally, the potential leader $p_i$ assembles a \textsf{proposal} message (\prm) in the form of  $PR_i^r(B^*, h, c, lc)$ containing:
1) the proposed block $B^*$, 
2) $B^*$'s height $h$, 
3) the proposal certificate $c$, 
and 4) the leader proof $lc$ (defined in Section~\ref{leader_election}).  
Then $p_i$ signs the message with her private key.  
Everyone can easily verify the validity of the \prm by checking the included block, signatures and certificates.

\subsection{Stage I: Block Proposal Broadcast}
\label{stageI}

After the leader selection step, \pname enters Stage I: block proposal dissemination.  
The objective of this stage is to propagate blocks proposed by all potential leaders to as many honest players as possible.

We use the well-known gossip protocol~\cite{terry1988ppidemic} to disseminate messages on the application layer overlay network formed by the same set of players.  A player sends/forwards a message to $m$ randomly selected players in each hop.  Parameter $m$ is called the \emph{fanout} and determines how fast the message propagates. 

Potential leaders initiate the gossip of their \prms.  
Each honest player, upon receiving a \prm, will first check its validity, and then forward all valid \prms. Note that there can be more than one valid block proposed in the same round, either because there are multiple potential leaders, or because a malicious leader can propose multiple valid blocks.  Players forward all valid blocks in this stage and leave the conflict resolution to Stage II.  

At the end of Stage I (after a fixed time interval $T_1$), we expect that most players have seen all block proposals in the round, assuming everything goes well.  Nevertheless, Stage II is able to handle all complicated situations.


\subsection{Stage II: Signature Collection}
\label{stageII}

The objective of Stage II is to disseminate signed messages
of players' votes for the block proposals, in the hope that honest players can commit a single valid block.  
Same as Stage I, we use gossip to propagate all messages.

\para{Message types. }
There are two message types involved in this stage.
Players can collectively sign a message by appending their signatures.
We say a message $M_X^r$ is \emph{$k$-signed} if $X$ contains \emph{at least} $k$ distinct players' signatures.

\noindent\textbf{$P$ message.} A \textsf{prepare} message ($P$ message) digitally signs a block.
Specifically, $P_i^r(B^r, h)$ means player $p_i$ signs her vote for the block $B^r$ at height  $h$ proposed in round $r$. In short, we say $p_i$ prepares or ``$P$''s $B^r$.

\noindent\textbf{$TC$ message.} A \textsf{tentatively-commit} message ($TC$ message) signs a proof of a $(2f+1)$-signed $P$ message. 
Specifically, $TC_i^{r}(B^r, h)$ proves that at least $2f+1$ players (including $p_i$) have $P$ed the block $B^r$ at height $h$ in round $r$.
In short, we say $p_i$ tentatively-commits or ``$TC$''s $B^r$.


\para{Stage II protocol. }
Algorithm~\ref{algo_stageII} outlines the expected behavior of an honest player $p_i$ in Stage II.  We model each $p_i$ as a finite state machine with local states listed at the beginning of Algorithm~\ref{algo_stageII}.  It performs actions based on current local state and the incoming messages.  

Lines~\ref{step1s} to~\ref{step1e}
describe the initialization procedure, in which $p_i$ checks all block proposals she receives in Stage I (by calling the function \emph{DecideMsg} in Algorithm~\ref{algo_func}).  
If $p_i$ received valid proposals in Stage I, she needs to decide which block to prepare (function \emph{DecidePMsg} in Algorithm~\ref{algo_func}).
In general, $p_i$ prefers a block proposal with larger proposal round, as it indicates a more recent block (lines~\algref{algo_func}{decidepmsg_s} to~\algref{algo_func}{decidepmsg_e}).
Finally, $p_i$ chooses exactly one block $\bar B$ for height $h$, and $P$s it (line~\algref{algo_stageII}{only_one_b} and line~\algref{algo_stageII}{p_the_only_b}, respectively). 

After initialization, the state machine of $p_i$ starts to handle incoming messages.  
Lines~\ref{step2s} to~\ref{step2e} in Algorithm~\ref{algo_stageII} outlines handler routines for these three different message types.  
$p_i$ only signs messages about the same block that she has $P$ed (line~\algref{algo_stageII}{sign_P} and line \algref{algo_stageII}{sign_TC}),
she can only $TC$ a block $B$ after she collects at least $2f+1$ 
signatures from the $P$ message about $B$, 
and she can only commit a block $B$ after she collects at least $2f+1$ signatures from the $TC$ messages about $B$ (line~\algref{algo_stageII}{can_tc} and~\algref{algo_stageII}{can_c}).
These rules ensure the safety of \pname.



\begin{algorithm}[tb]
\caption{Stage II workflow for each player $p_i$. }
\label{algo_stageII}

\small

 \algrenewcommand\algorithmicwhile{\textbf{On}}
  \algrenewcommand\algorithmicdo{\textbf{Do}}
   
%

\textbf{Constants:}

-- $\mathbf G$: the consensus group of N players

-- $r$: the current round number 

\textbf{State Variables:}

-- $s_i$: $p_i$'s local state, i.e., $\langle B_{root}, h_{root}, B_{tc}, F \rangle$

-- $S$: the set of all valid proposals received in Stage I

\begin{algorithmic}[1]
	\label{main}
	
		\State $phase \gets \textsf{Init}$ 
		\label{step1s}
		\State $msg \gets$ \Call{DecideMsg}{} \Comment See Algorithm~\ref{algo_func}
		\If{$msg \neq null$}
			\State $P(\overline B, h) \gets msg$ 
			\Comment $\overline B$ is the block $p_i$ votes for
			\label{only_one_b}
			\State $phase \gets \textsf{Ped}$
			\State $X_P \gets \{ i \}$  \Comment The players that have $P$ed $\overline B$ 
			\State Prepare $\overline B$ by gossiping $msg$
			\label{p_the_only_b}
		\EndIf
	\label{step1e}

	\vspace{2 mm}
	
	\While{receiving a valid $P_{X'}^r(B, h)$ message}
	\label{step2s}
		\If{ $phase = \textsf{Ped}$ \textbf{and} $B=\overline B$}
		\label{sign_P}
			\State $X_P \gets X_P \cup X'$
			\State $sig_{P} \gets$ signatures of players in $X_P$\footnotemark
			\State Sign $P_{X_P}^r(B, h)$ with $sig_P$
			\If{$P_{X_P}^r$ is (2f+1)-signed}
				\label{can_tc}
				\State $phase \gets \textsf{TCed}$
				\State $X_{TC} \gets \{ i \}$ \Comment The players that have $TC$ed $\overline B$ 
				\State $s_i \gets \langle B_{root}, h_{root}, B, r \rangle$ 
				\label{tc_action}
				\State Tentatively commit $B$ by gossiping $TC_i^r(B, h)$
			\Else
				\State Forward the signed message $P_{X_P}^r$ with gossip
			\EndIf
		\EndIf
	\EndWhile
	
	\vspace{2 mm}
	
	\While{receiving a valid $TC_{X'}^r(B, h)$ message}
		\If{$phase \neq \textsf{Init}$ \textbf{and} $B=\overline B$}
			\label{sign_TC}
			
			\State $X_{TC} \gets X_{TC} \cup X'$
			\State $sig_{TC}\gets$ signatures of players in $X_{TC}$
			\State Sign $TC_{X_{TC}}^r(B, h)$ with $sig_{TC}$

			\If{$phase \neq \textsf{Ced}$ 
					\textbf{and} $TC_{X_{TC}}^r$ is (2f+1)-signed}
				\label{can_c}
				\State $s_i \gets \langle B, h, \epsilon, 0\rangle$ 
				\State Commit $B$ on the local blockchain
				\State $phase \gets \textsf{Ced}$
			\EndIf
			\State Forward the signed message $TC_{X_{TC}}^r$ with gossip
		\EndIf
	\EndWhile
	
	

\label{step2e}
\end{algorithmic}
\end{algorithm}

\begin{algorithm}
\caption{Deciding which block to prepare in Stage II.}
\label{algo_func}
\small
\begin{algorithmic}[1]
\Function{DecideMsg}{} 
\label{decidemsg_s}
\If {$S = \phi$}  \Comment{Received no valid proposals}
		\State \Return null
		\label{decide_tc_e}
\Else{} 
	\Comment Received one or more valid proposals
	\State \Return \Call{DecidePMsg}{}
\EndIf
\EndFunction

\vspace{2 mm}

\Function{DecidePMsg}{}
	\State $r_p^* \gets \max_{B \in S} r_p(B)$
	\State $\overline B \gets \arg\min_{B_j \in S, r_p(B_j) = r_p^*}  L^r(j)$\footnotemark
	\State Denote by $PR(\overline B, h, c)$ the proposal message about $\overline B$
	\If { $h=h_{root} +1$ }
		\If {$B_{tc} = \epsilon$ }
			\State $s_i \gets \langle B_{root}, h_{root}, \epsilon, 0 \rangle$ 
			\State \Return $P_i^r(\overline B, h)$
		\Else{}
			\label{decidepmsg_s}
			\label{tced_s}
			\If {$r_p(\overline B) > F$}	
				\State $s_i \gets \langle B_{root}, h_{root}, \epsilon, 0 \rangle$ 
				\State \Return $P_i^r(\overline B, h)$
			\ElsIf {$\exists B' \in S$ s.t. $B' = B_{tc}$ \textbf{and} $r_p(B') \geq F$\footnotemark}
				\State $s_i \gets \langle B_{root}, h_{root}, B', r_p(B')\rangle$ 	
				\State \Return $P_i^r(B', h)$
			\EndIf	
			\label{tced_e}
		\EndIf
	\EndIf
\Return null

\label{decidepmsg_e}
\EndFunction

\end{algorithmic}
\end{algorithm}

\subsection{Reducing Signature Sizes}
\label{cosig}

\pname protocol requires signatures from over 2/3 of the players.  
To reduce the storage and communication overhead of signatures,
we adopt the techniques in~\cite{boneh2001short,boneh2003aggregate} to aggregate these signatures into a compact \emph{multi-signature} form.




\footnotetext[7]{Other players' signatures are in the received messages.}
\footnotetext[8]{Among all blocks with the largest proposal round in $S$,
$\overline B$ is the block whose proposer has the smallest leader score.}
\footnotetext[9]{Note that $r_p(B')$ is actually the proposal round of the proposal message about $B'$, which is not necessarily equal to $r_p(B_{tc})$.}

The cryptographic signature of a player $p_i$ involves a hash function $H$, a generator $G$, a private key $x_i$, and a public key $V_i=G^{x_i}$. A player holding the private key $x_i$ can sign a message $M$ by computing $S_i=H(M)^{x_i}$, and others can verify it by checking whether $e(G, S_i)$ is equal to $e(V_i, H(M))$ with a given bilinear map $e$. 
To track which signatures we have received,
we append an integer array $n$ of size $N$ to the signature, and by signing a message $M$, a player computes $S_i=H(M)^{x_i}$, and increments the $i$-th element of $n_{S_i}$. The combination is the signature for aggregation, and we denote this process by $sign_i(M)=(S_i,n_{S_i})$. 

An important property of the aggregated signature is that
we can put in new signatures in an arbitrary order, avoiding the risk of adaptive chosen-player attack that Byzcoin\cite{kogias2016enhancing} faces.  
Aggregating signatures is simply multiplying the BLS signature and adding up the array $n$.  Thus, the aggregated signature (aka \emph{multi-signature}) is $S=H(M)^{\sum_i x_i\cdot n_S[i]}$. 
We denote the process by $aggregate(S_1, S_2,...)=(S,n_S)$. 
Let $(S_1, n_{S_1})$ and $(S_2, n_{S_2})$ be two multi-signatures, 
we can combine them by computing  $aggregate(S_1, S_2) = (S_1 * S_2, n_{S_1} + n_{S_2})$.  The array $n$ tracks who have signed the message.  
Everyone can verify the multi-signature by checking whether $e(G, S) = e(\prod_i V_i^{n_S[i]},H(M))$. 

\cite{micali2001accountable,boneh2003aggregate} points out that aggregating signatures of the same message can be vulnerable to chosen-public-key attack. This attack can be avoided if the participates can prove they have the private key to their announced public key, either forced by a trusted third party or by a zero-knowledge-proof bootstrap process proposed by \cite{micali2001accountable}. We choose this method because it's acceptable with the help of PKI. 

Another method proposed in \cite{boneh2003aggregate} computes $H(M+V_I)$ instead of $H(M)$ so each player signs different messages. 
Since everyone knows each others' public keys, the result is still verifiable without increasing the data size. This method does not involved a trusted third party or online bootstrap process, but it forces the algorithm to compute the bilinear map $N$ times, instead of one time when the messages are the same. It can cause the verification time 100 times slower, and thus can only be adopted when the number of players in a system is small (less than 200).

The signature aggregation process significantly reduces memory utilization.
Although the multi-signature still has size $O(N)$ asymptotically, a 4-byte integer is enough for each element in $n_S$ in most cases. 
With 1,000 players using a 2048-bit signature, naively it takes 256 KB to store these signatures, but with aggregation, it requires only 4256 bytes, or 1/60 of the original size.
The optimization is more efficient as the system scales larger. 
For the case when the number of signers is small, the array is sparse and thus easily compressible.


\subsection{Player Recovery from Temporary Failures}
\label{block_inquiry}

In normal cases, blocks and signatures are broadcast to all players. If a player misses a block in round $r$ due to temporary failures, it can catch up in subsequent rounds using the following (offline) recovery procedures:

If player $p_i$ receives a valid signature of enough signers in round $r$ but fails to receive the block itself, $p_i$ will check the signature and try to contact someone who has signed the block to retrieve it. 

If player $p_i$ recovers from an extended crash period and/or data loss, it should try to retrieve all lost blocks and proofs. 
She can only continue participating in the protocol after she recovers the entire history.  
As the committed blocks are offline-verifiable, blocks with valid signatures from any player are sufficient for recovery.


\subsection{Security Analysis}
\label{security_analysis}

We can prove that 
\pname provides safety (as defined in Section~\ref{problem_definition}) in fully asynchronous networks.  Adding \emph{partial synchrony} assumption, it also achieves liveness. 
We only list some key lemmas here
and leave the complete proofs in Appendix~\ref{safety} and \ref{liveness}.

\begin{myLem}
	\label{no_tc}
	If an honest player $p_i$ commits a block $B$ at height $h$ in round $r$,
	no player will ever $TC$ any other block $B'$ at any height $h' \leq h$ in any later rounds.
\end{myLem}


\para{Proof sketch.}
At least $f+1$ honest players will not $P$ any blocks whose proposal
rounds are no larger than $r$ (line~\ref{tced_s} to~\ref{tced_e} in Alg.~\ref{algo_func}). 
Therefore, at least $f+1$ honest players will not $P$ any other block at height $h' \leq h$ after round $r$, proving Lemma~\ref{no_tc}. And Lemma~\ref{no_tc} leads to safety, because no block can be committed without honest players signing $TC$ messages.

The following two lemmas prove the liveness under the partial synchrony assumption. 
\begin{myLem}
\label{liveness_p}
If in round $r$, for any honest player $p_i$ we have $s_i = \langle B_{root}, h_{root}, \epsilon, 0 \rangle$, then there exists a round $r' > r$ and an honest player $p_j$ such that $p_j$ $P$s some block at height $h = h_{root}+1$ in round $r'$.
\end{myLem}

\begin{myLem}
\label{liveness_c}
If in round $r$, there exists some honest player $p_i$ with state $s_i = \langle B_{root}, h_{root}, B_{tc}, F\rangle$ $(B_{tc} \neq \epsilon)$,
then there exists a round $r' > r$ and an honest player $p_j$ such that $p_j$ commits a block at height $h = h_{root}+1$ in round $r'$.
\end{myLem}

%
%

\para{Attacks beyond the protocol layer.}
In addition to the adaptive chosen-player attacks and other attacks causing communication problems, an adversary can design attacks on the system implementations, including:
1) \emph{computation resource saturation} attack, 
where the adversary may disseminate a large number of invalid messages to
the honest players, consuming their CPU cycles for useless signature verification,
and 2) \emph{signature counter overflow} attack, 
where the adversary may craft valid multi-signatures where some counters
are close to the maximum integer, and thus careless signature aggregating of honest players may cause an integer overflow resulting in incorrect signatures.

%

Note that both attacks can only cause liveness problems, rather than correctness problem. 
We will describe our countermeasures to both attacks in Section~\ref{implementation}.

\section{Implementation-level Optimization}
\label{implementation}

As we mentioned, \pname combines protocol level design and implementation level optimizations to achieve high performance.  In this section, we introduce the important optimizations. 



\para{Asynchronous transaction dissemination and hash-only blocks.} 
In our protocol, messages only contain \emph{hashes} of the transactions to reduce message size.  
Raw transactions are gossiped among all players asynchronously, independent of protocol stages.
In the case that a player does not have the raw transaction data when she receives the blocks, she retrieves the transactions from others before she can process the block. In practice, the gossip protocol often does a good job replicating the transactions, and thus this retrieval finishes fast.  


\para{Continuous gossiping in Stage II voting.  }
As we need all honest players to receive proposed blocks and others' votes to achieve liveness, we do not limit the fanout.  Instead, each player continuously sends block or $P$/$TC$ messages to random neighbors until the end of the stage.  
However, we do put on a limit of concurrent connections to avoid overloading any player, which we set to 5 by default in our implementation. Section~\ref{scalability} provides a detailed analysis of the fanout limit. 

\para{Blacklisting obvious problematic players.  }
While there is no way to ensure message delivery, if a player detects an obvious communication problem (connection failure, timeout etc.) with a peer, she will ``blacklist'' the peer (i.e. stop sending to it) for a short time period $T_o$ (typically half a round time).  On subsequent failure with the same peer, she will additively increase $T_o$, until she receives a message from that peer, or successfully retries.  This backoff mechanism effectively limits the wasted attempts to connect to failed nodes.  



\para{LIFO processing stack.  }
In Stage II, each node can concurrently receive multiple messages with signatures for processing (verification + aggregation).  Sometimes the messages arrive faster than the server can process them.  We put them in a last-in-first-out (LIFO) \emph{stack}, instead of a queue.  This is because it is likely that later arriving messages contain more signatures, or sometimes even a super-set of signatures in earlier messages.   


\para{Preventing signature overflow. } 
In the aggregated signature described in Section~\ref{cosig}, we have the array $n$ with $N$ $B$-bit integers (we have $B=32$ as default). An adversary can craft a valid signature so that the element corresponding to her signature is $2^B-1$.  This attack prevents honest players who have this signature from further aggregating the signature array because otherwise, the element will overflow.  

We prevent such attack by restricting the growth of the maximum element in $n$, $max$, based on the number of signers $s$. On receiving a new message, the player tries to aggregate it into her local signature array, and check if the result satisfies $max \leq s$ or $\log_2(max)< B*s/N$. If so, the player updates her local array; otherwise, she drops the incoming message. 

\section{Evaluation}
\label{evaluation}
\setlength{\abovecaptionskip}{0pt plus 3pt minus 2pt}
\setlength{\belowcaptionskip}{0pt plus 3pt minus 2pt}
%
%
%

We evaluate the performance of \pname using both simulations and real testbed experiments.

\subsection{Evaluation Setup}

\para{\pname prototype implementation.  } We implement the \pname prototype in Java.  We use \texttt{pbc}~\cite{lynn2007implementation} library (with \texttt{JPBC}~\cite{de2011jpbc} wrapper) for cryptographic computation and use \texttt{grpc-java}~\cite{grpc-java} for network communication. As for signature parameters, we choose the default \emph{a}-type parameter provided by \texttt{JPBC}~\cite{jpbc-parameter}, and use it to generate 1024-bit BLS signature.  The entire system contains about 5,000 lines of Java code excluding comments. 


\para{Testbed. }  We build a testbed with up to 140 \texttt{t2.medium} instances evenly distributed on Amazon EC2's all 14 regions on 5 continents. 
We experimentally measure the network condition between the instances.  Within a region, we have less than 1ms latency and about 100 MBps bandwidth, and latencies across regions are hundreds of milliseconds with the bandwidth varying from 2 MBps to 30 MBps.  We believe the testbed is a good emulation of a typical multi-datacenter WAN.  



Each instance acts both as client and server in the system.  As the client, it generates transactions with an exponentially distributed inter-arrival time.  Each transaction is 250 bytes, a typical Bitcoin transaction size (and used in evaluations of \cite{miller2016honey} too).  These transactions are submitted to the servers on the same instance. 


\para{Simulation for larger scales.  }
Limited by the testbed scale, we depend on simulations to analyze larger scale behavior of our signature collection process.  We set the network latency to an exponential distribution with a mean of 300 ms, a typical value for today's Internet~\cite{ledlie2007network}, and set the bandwidth to 500~KBps, a generous estimation after subtracting the bandwidth consumed by constantly gossiped transactions. 
We set the packet loss rate to $1\%$ across all links (higher than many Internet links). In the simulator, we do not actually verify signatures, but set the signature verification time to be consistent with the performance we get on AWS \texttt{t2.medium} instances.\footnote{The verification time consists of an 11 ms constant overhead for computing bilinear map functions and another $0.11k$ ms for $k$ signers. That means 12.1ms for 10 signers and 1111ms for 10000 signers. }
 All faulty players in testbed experiments and simulation simply fail by crashing. 


\begin{figure*}[htb]
\subfigure[Latency with varying workload.]{
	\includegraphics [width=0.33\textwidth]{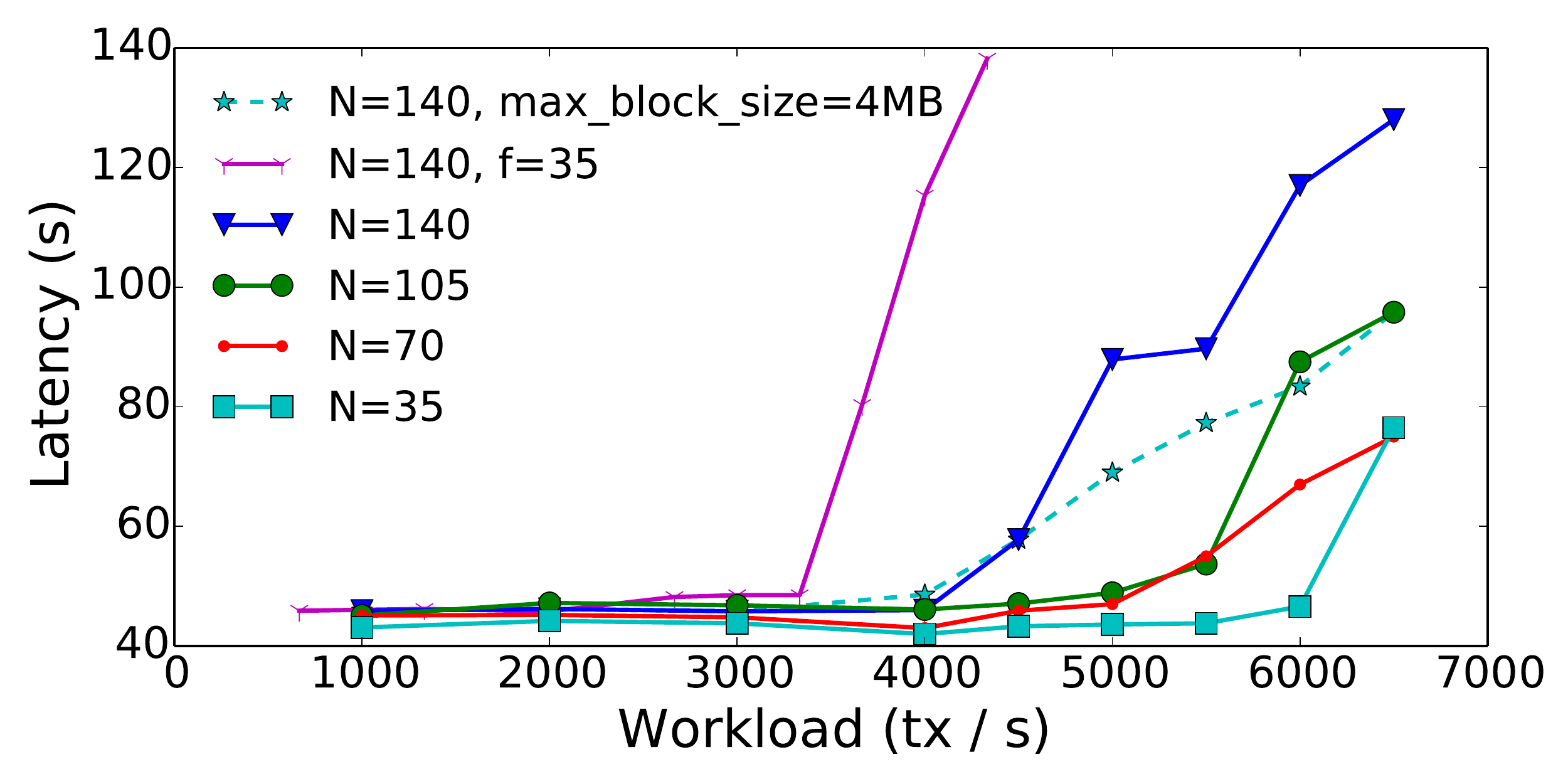}
	\label{fig:latency_workload}
}
\subfigure[Throughput with varying workload]{
	\includegraphics [width=0.33\textwidth]{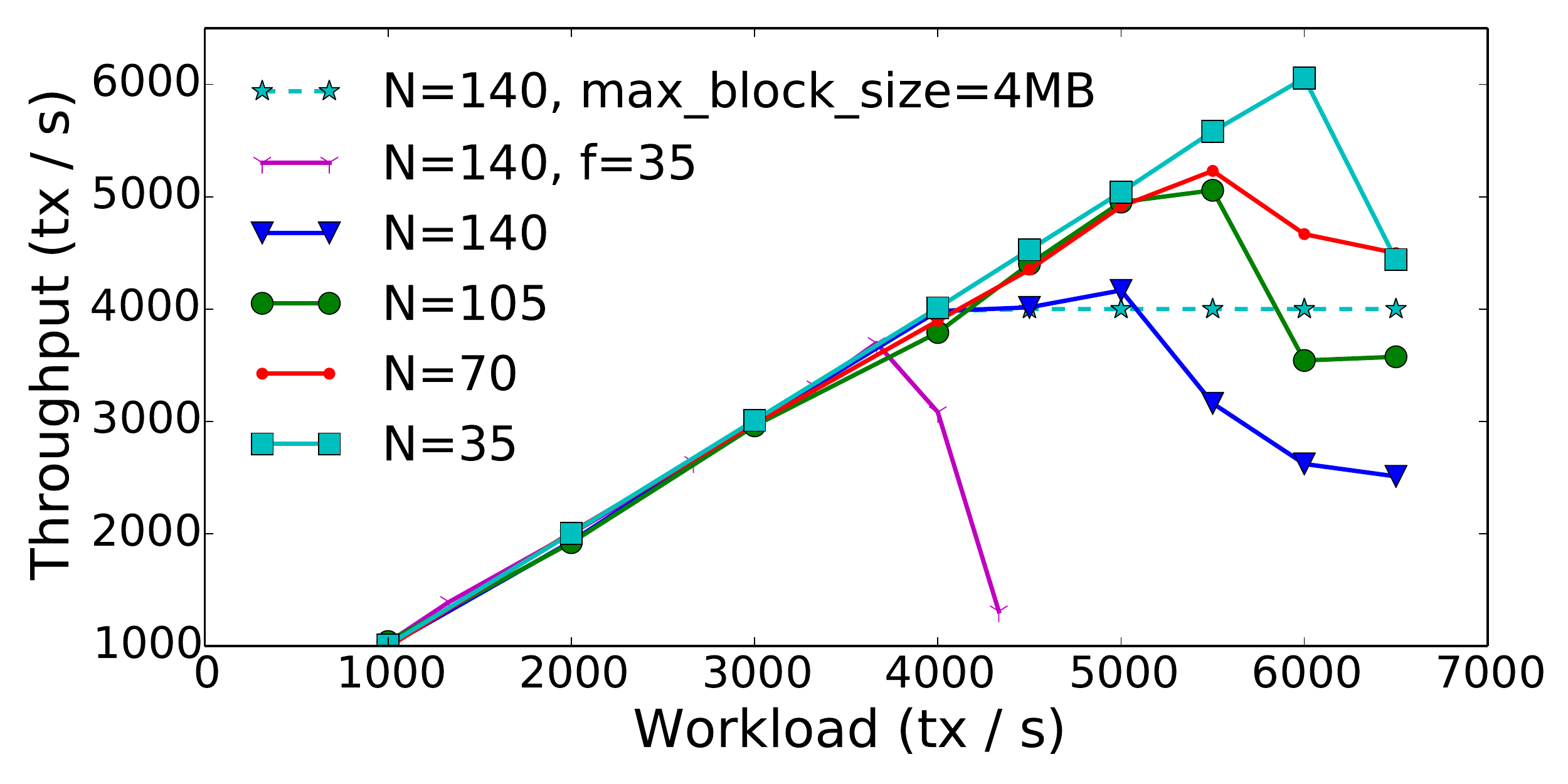}
	\label{fig:throughput_workload}
}
\subfigure[Latency vs. throughput]{
\includegraphics [width=0.33\textwidth]{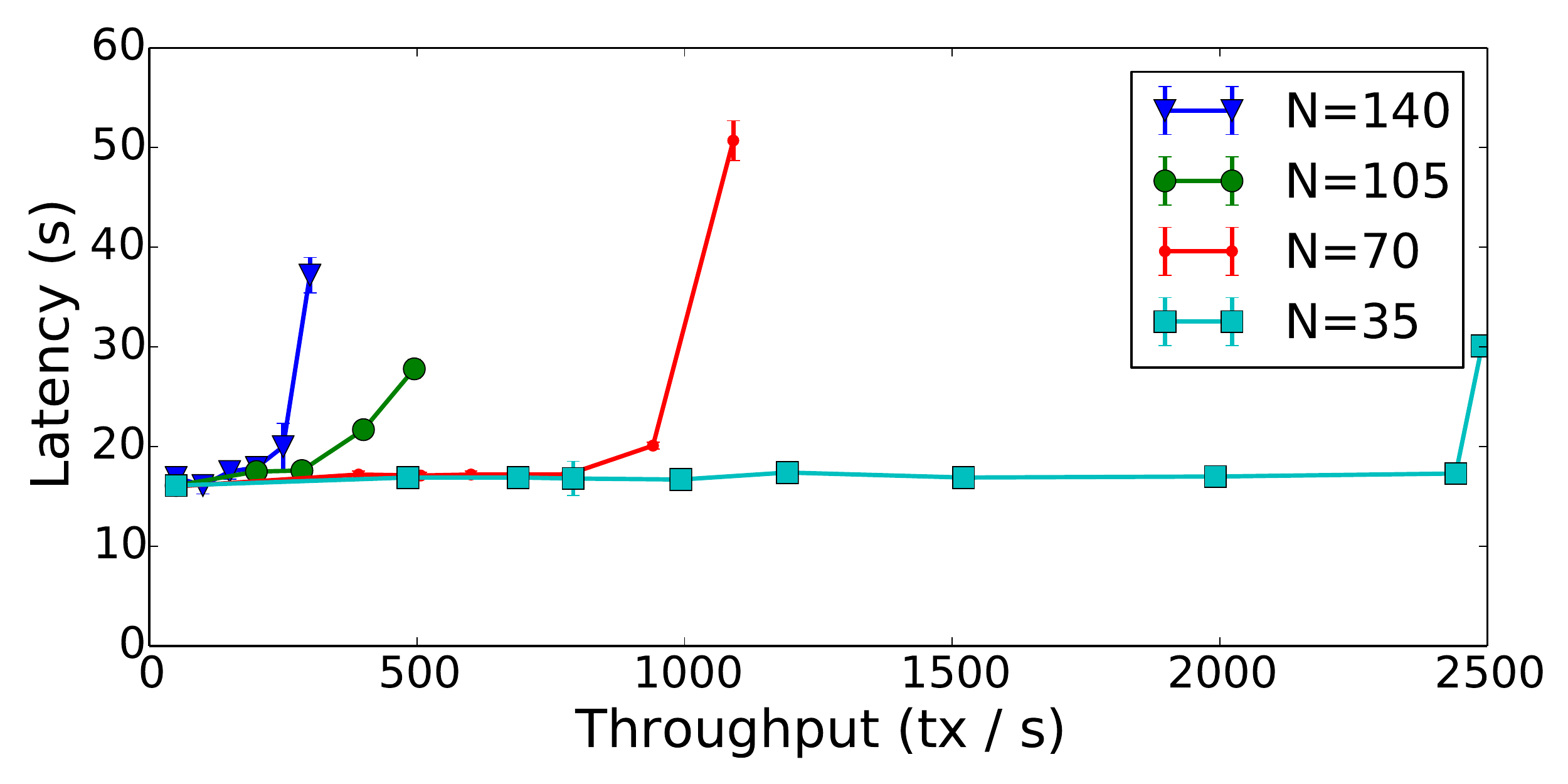}
\label{fig:low_latency}
}
\caption{Performance under different configurations. (a) and (b) are throughput-optimized. (c) is latency-optimized.}
\end{figure*}

\para{Key Configuration parameters.  }
\label{configuration}
There are several configuration parameters to tune.  The most important ones include the round time $T$ and maximum block size \mbs. Both are correlated and affect system scalability and performance significantly.  We discuss their impacts in Section~\ref{parameter}.  

We set $T = 30$ sec (25 seconds for Stage I and 5 seconds for Stage II), and $\mbs=8$MB.  Recall that in \pname we only propagate transaction hashes (and send actual transactions asynchronously).  Given that each transaction hash is 256 bit, a block can contain at most 250K transactions.  
With an average transaction size of 250 bytes, the corresponding raw transaction data for a block is about 62.5MB.

\subsection{Real Testbed Performance}

Here we present the performance metrics on the EC2-based testbed.  We test two configuration settings, one optimized for throughput (the default setting), and the other optimized for transaction commit latency.  

\subsubsection{Throughput-optimized configuration}
\label{throughput}
Using default parameter settings (Section~\ref{configuration}), we run experiments on 35, 70, 105, 140 instances for 1200 seconds each using different workload from 1,000 tps to 7,000 tps.  
Figure~\ref{fig:throughput_workload} and \ref{fig:latency_workload} plot the throughput and average commit latency, respectively.  We have the following observations:

1) Without overloading the system, the average commit latency is a little over 40 seconds.  This is consistent with our theoretically expected latency of 1.5 rounds.  

2) With 35 players, we can sustain a 6,000 tps workload.  With 140 players, we can still support 4,000 tps.  Comparing to the reported numbers in HoneyBadgerBFT~\cite{miller2016honey} (using similar EC2 testbeds, but fewer geo-locations), we double the throughput and reduce the latency by 80\% with even more players and more regions.  

3) When the system gets overloaded, the throughput actually \emph{drops}. 
This is because, with a 30-second round time, there is not enough time to propagate all blocks to everyone, causing incomplete rounds and thus reducing the effective throughput (aka. goodput).  To prevent such situation, we limit the $\mbs$ as an admission control mechanism, just like most blockchains do.  In fact, the dashed line in Figure~\ref{fig:throughput_workload} shows that when limiting $\mbs$ to 4MB (i.e. 125K transaction per block, or 4167 tps), we can sustain the maximum throughput even on overloading.  Of course, overloading still causes the latency to go up, but there is no difference from any queuing system. 

4) \pname tolerates failures quite well with small overhead. As Figure~\ref{fig:latency_workload} and \ref{fig:throughput_workload} shows, 35 faulty ones among 140 total nodes show little influence on the system's throughput or latency without overloading.  The only impact of these failures is decreasing the maximal throughput by about 10\%, from 4,000 tps to 3,600 tps.

\subsubsection{Latency-optimized configuration}
\label{latency}

The default setup uses large block sizes and long round time (30 seconds) to improve overall throughput.  For applications that are more latency-sensitive, we provide an alternative configuration.   
We reduce the round time $T$ to 10 seconds with 5 seconds for each stage.  We also disable block-existence probing (see Section~\ref{implementation}) to further reduce latency. Then we repeat the same set of experiments, and Figure~\ref{fig:low_latency} shows the latency we can achieve under different workloads. 

Like the previous case, we can get less-than-17-second latency and stable throughput until overloading.  We can sustain over 200 tps with 140 nodes, 600+ tps with 70 nodes or 2,400+ tps with 35 nodes.  However, the 10-second round offers very tight time budget for blocks to propagate.  Larger blocks have little chance to complete propagation, causing the latency to go up quickly on overloading.  Thus, a carefully controlled block size is even more essential. 

In comparison, HoneyBadgerBFT cannot offer a low latency configuration in a relatively large group because of its $O(N^2)$-message-per-node complexity.  Evaluation in~\cite{miller2016honey} shows that the latency with only 64 players cannot even go below 200 seconds.  




\subsubsection{Latency breakdown }

\begin{figure}[tb]
\includegraphics [width=0.45\textwidth]{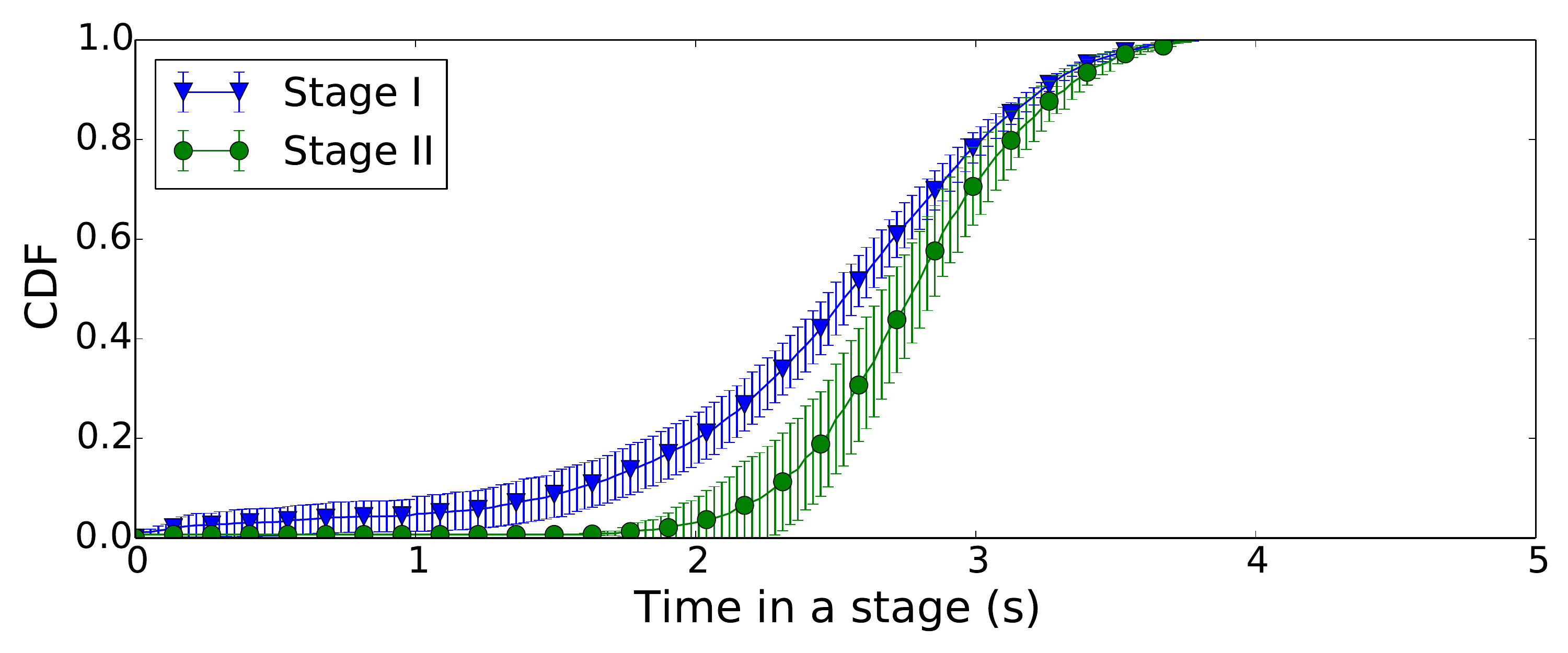}
\caption{CDF for each player's stage completion times, using latency-optimized configuration. The error bars are 95\% confidence intervals.}
\label{fig:sig_time}
\includegraphics [width=0.45\textwidth]{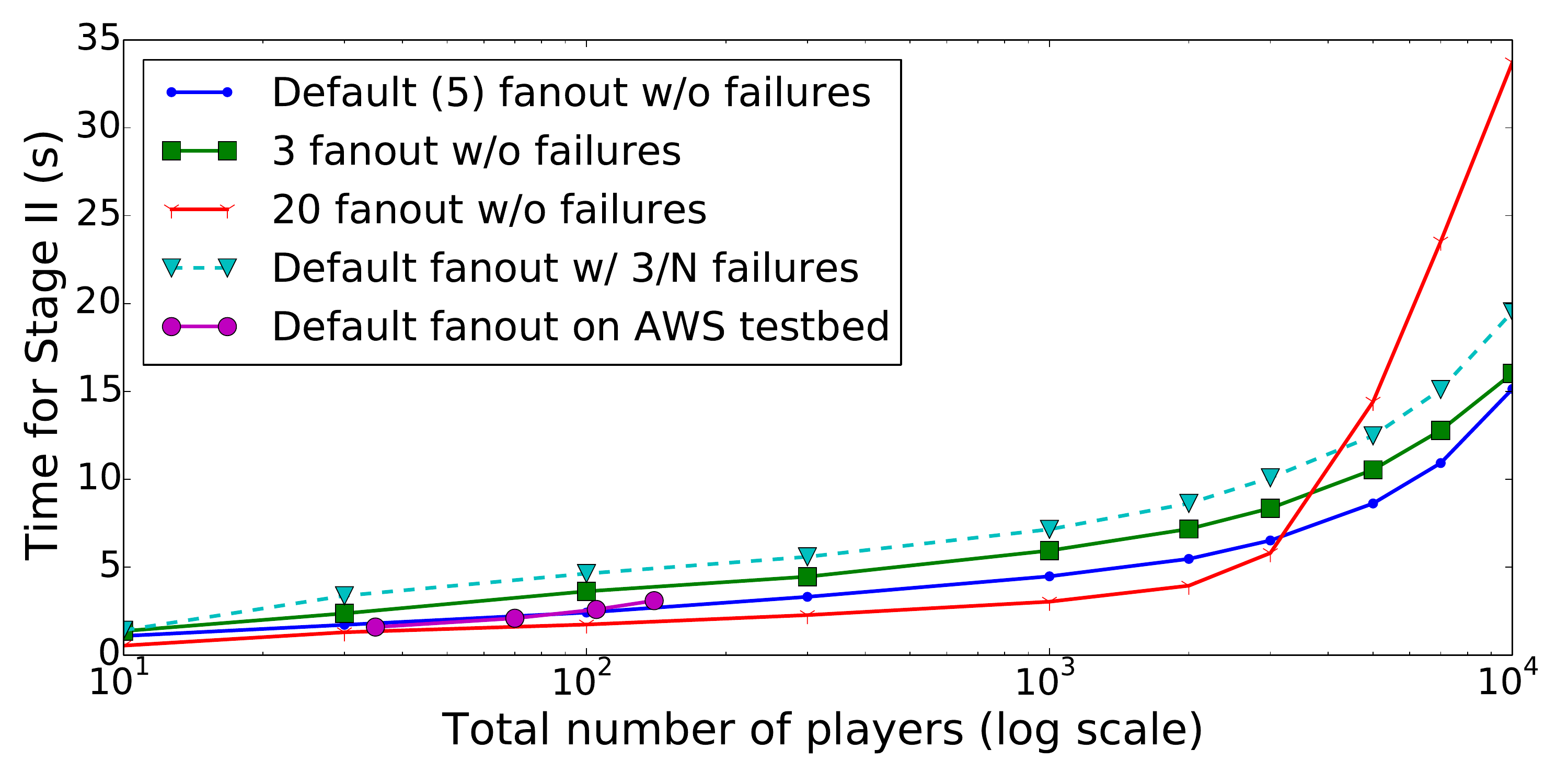}
\caption{Simulation results for Stage II completion time.}
\label{fig:scalability_simulation}
\end{figure}

While the transaction commit latency is largely determined by the round time, we want to take a closer look at how fast a player can commit a block within a round.  We plot the cumulative distribution (CDF) of the time taken for players to commit a block, using the same low-latency configuration with 140 nodes and 200 tps workload.  Figure~\ref{fig:sig_time} shows the CDF of completion time for both stages on each node.

We can see that Stage II is only slightly slower than Stage I, especially at the slowest player (both at about 4 seconds).  It is a little counter-intuitive as Stage II consists of 2 rounds of messaging (P and TC messages) vs. Stage I has only one. The reason is that to complete Stage I, every player has to receive the block from all leaders, in order to determine the least \emph{leader score}.  In comparison, Stage II only need votes from any $2f+1$ players. 


\subsection{Scalability}
\label{scalability}

Limited by the resources on the testbed, we evaluate the scalability on systems larger than 140 nodes using simulation.  In the simulation, we focus on the completion time of Stage II, because it is the core and most complicated part of \pname, while the performance of Stage I is no different from other gossip-based systems.  

Using the default settings, we show the time required to complete stage II with 10 to 10,000 players, i.e., all honest players receives a $TC$ multi-signature signed by more than 2/3 players.  Figure~\ref{fig:scalability_simulation} shows the results using different combinations of failure modes and optimizations.  The key observations are:

1) To calibrate the simulator, we also reproduce the testbed results (for up to 140 nodes) in Figure~\ref{fig:scalability_simulation}.  We can see that it fits our simulation results quite well.  


2) \pname scales well.  Even with 10,000 players, we can still finish Stage II within 15 seconds only.  
This is a direct benefit of using gossip-based application overlay network and asynchronous multi-signature, which fully exploits the redundant paths while keeping the bandwidth consumption small. The time grows faster when the number of players $N$ is large, because the overhead for signature verification increases linearly with $N$. But this overhead only becomes significant after the total number goes beyond 3000, and can be reduces by stronger hardwares.


3) With 10000 players and 1/3 of them being faulty by crashing, Stage II completion time slowdowns from 14.97 seconds to 19.53 seconds.
The robustness of the protocol comes from the gossip mechanism and the order-independent signature aggregation algorithm. 

4) A small gossip fanout, i.e., the number of outbound connections, can fit most environments. A large fanout (like 20 in Figure~\ref{fig:scalability_simulation}) will saturate the network and cause higher latency due to queueing effects. Although a small fanout may not fully utilize the network when the system has fewer players, the cost is not significant since most time of a round is allocated to stage I. 


%


\subsection{Configuration Parameters}
\label{parameter}

As we have seen in Section~\ref{throughput} and~\ref{latency}, \mbs and round time $T$ affect system performance significantly.

With $N$ players, the \mbs is proportional to three parameters~\cite{croman2016scaling,terry1988ppidemic}: 1) $1 \over \log N$, 2) round time $T$, and 3) the network bandwidth.  
That means, in order to increase the number of nodes from 100 to 10,000, we need to either decrease the \mbs by half or double the round time $T$ given a fixed bandwidth.  

In the remaining of the section, we experimentally evaluate their impacts using all 140 nodes in the testbed. 

\para{Max block size. }
\label{blocksize}
As we have discussed, the parameter \mbs serves as an admission control mechanism to avoid overloading the system.  

\begin{figure}[tb]
\center
\includegraphics [width=0.45\textwidth]{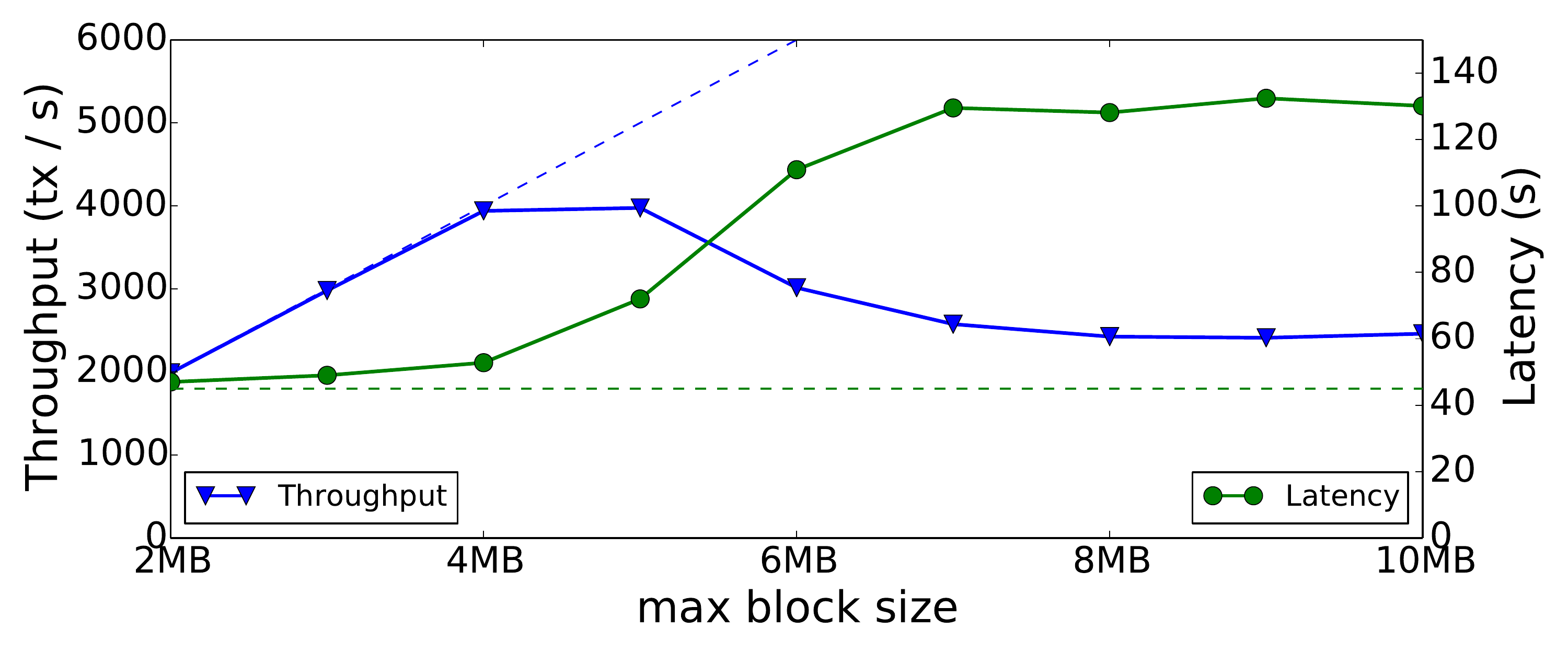}
\caption{Influence of \mbs on the performance.}
\label{fig:maxtps}
\includegraphics [width=0.45\textwidth]{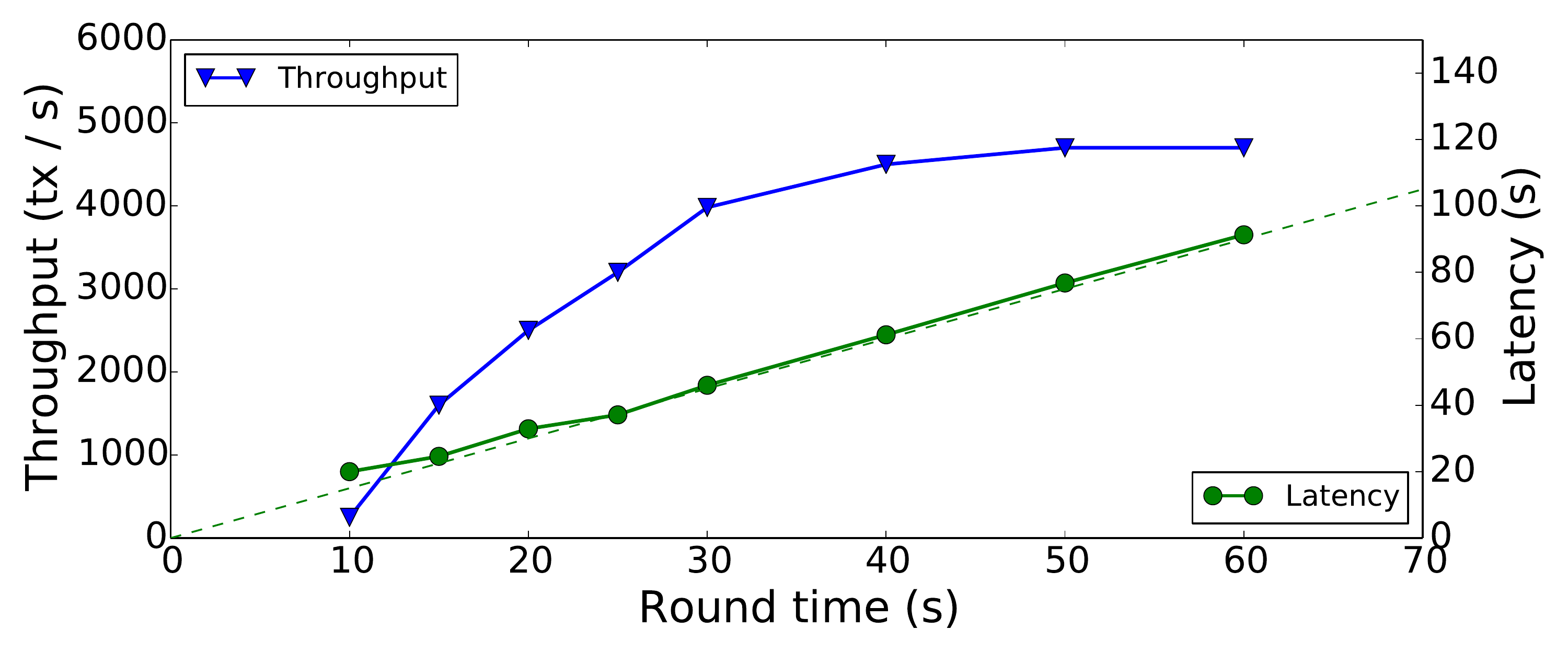}
\caption{Influence of round time on the performance.}
\label{fig:roundtime}
\end{figure}

Keeping $N=140$ and $T=30$s, we vary \mbs from 2MB to 10MB, corresponding to 60K to 300K transactions per block, or 2K to 10K tps.  In each round, we generate workload that equals the \mbs.   Figure~\ref{fig:maxtps} plots the results.  The dashed line in Figure~\ref{fig:maxtps} shows the ideal case where the system has infinite capacity.

The actual throughput of the system is around 4,000 tps, as we presented in Section~\ref{throughput}.  We can see that for a \mbs smaller than 4MB, the actual throughput increases with the \mbs and roughly follows the ideal line.  At around 4,000 tps, the system saturates.  If the \mbs is significantly larger than what the system can handle, the throughput decreases because some rounds will end before the players can fully propagate the blocks. 


\para{Round time $T$ vs. throughput. }
\label{roundtime}
Given a round time $T$, it is easy to calculate the expected transaction commit latency when the system is under a normal workload, which is $1.5T$.  A latency significantly larger than this value indicates system overloading.  

Here we experimentally find the maximum throughput we can obtain under different $T$s, without overloading the system.  
Of course, because of the global clock synchronization error $\Delta$ and message propagation latency,  we require $T$ be at least 10 seconds.  

Figure~\ref{fig:roundtime} plots both the throughput and latency under different $T$ settings.  We verify that we are able to keep the latency very close to the expected latency of $1.5T$.  
We observe that choosing a small $T$ significantly reduces the maximum throughput.  The throughput even drops super-linearly as $T$ gets smaller than 30 seconds.  This is because when $T$ and \mbs are both small, the network setup overhead becomes non-negligible, further reducing the number of block transfers we can complete during the round.  Fortunately, a $T$ of 40 seconds already supports the max throughput in a 140-node system, still much faster than existing solutions.

\section{Conclusion and Future Work}

There are two types of approaches to build scalable blockchain systems: Some focus on theoretically provable protocols (e.g. HoneyBadgerBFT) even with high performance overhead, and others adopt best-effort hacks and hope it works most of the time (e.g. Bitcoin).  We believe there should be a middle ground: we insist on a system with provable safety under a very strong adversary assumption, while adopting the best-effort approaches to increase the probability of staying alive.  We use \pname to demonstrate such a system.  At the protocol level, using cryto-based secrete leader election and multi-round signature-based voting, we can guarantee safety, and by adding the partial synchrony assumption, we can also prove liveness.  At the same time, we adopt implementation-level techniques such as gossiping, failure discovery and asynchronous signature verification to increase the probability of liveness.  With evaluations on a real 140-server wide-area network, we show that \pname both scales well and provides high performance.   


As the next steps, we want to extend the protocol to allow players to join/exit.  We also want to explore the approaches of integrating \pname with other data-center-focused protocols to create a hybrid protocol, further improving its performance.  

%
%


{\footnotesize \bibliographystyle{acm}
\bibliography{ref} 
\appendix
\newtheorem{Lem}{Lemma}[section]
\newtheorem{Pro}{Proof}[section]
\newtheorem{The}{Theorem}

\section{Safety}
\label{safety}
\begin{Lem}
	\label{appendix_no_tc}
	If an honest player $p_i$ commits a block $B$ at height $h$ in round $r$,
	no player will ever $TC$ any other block $B'$ at any height $h' \leq h$ in any later rounds.
\end{Lem}

\para{Proof. }
An honest player $i$ commits a block $B$ with a height of $h$ means that at least 2f+1 players have $TC$ed $B$ in round $r$, so at least f+1 honest players have $TC$ed $B$ in round $r$. Those who commit the block $B$ successfully will reject all messages for blocks of a height no higher than $h$, so we only consider the case where they fail to commit the block $B$.

There is no other block proposed before round $r$ whose proposal round is larger than $r$. These f+1 honest players should have set their local $F=r$ according to line~\ref{tc_action} in Algorithm~\ref{algo_stageII}, so they will not $P$ any blocks proposed before round $r$. 

Also, these f+1 honest player will not send $TC$ message for any block of height less than $h$ after round $r$, because they have committed the same block of height $h-1$. Thus, no block of height less than $h$ can be committed after round $r$, meaning no one is able to propose a new valid block with a height of $h' \leq h$ and with a proposal round larger than $r$ after round $r$.

The only case left is that there can be more than one valid block proposed in round $r$, whose proposal rounds are the same with $B$'s.
In this case, by Algorithm~\ref{algo_stageII}, the $f+1$ honest players who have $P$ed $B$ in round $r$ will not $P$ any other round-$r$ block any more, so no other blocks proposed in round-$r$ can be $TC$ed, which requires $P$ messages from at least $f+1$ honest players.

To sum up, at least f+1 honest players will not $P$ any other block with a height of $h' \leq h$ after round $r$, and thus no one can $TC$ any other block with a height of $h' \leq h$ after round $r$, which requires at least 2f+1 $P$ messages.  $\hfill \square$

\begin{Lem}
	\label{appendix_no_c}
	If an honest player $p_i$ commits a block $B$ at height $h$ in round $r$,
	no player will ever commit any other block $B'$ at any height $h' \leq h$ in any later rounds.
\end{Lem}

\para{Proof. }
	By Lemma~\ref{appendix_no_tc}, we get that no one can $TC$ any new block with a height of $h' \leq h$ after round $r$, so no honest player can commit any block with a height of $h' \leq h$ after round $r$, as he cannot collect enough $TC$ messages.	$\hfill \square$

\begin{The}
	\pname protocol achieves safety.
\end{The}

\para{Proof. }
	Any committed block will be prepared by at least f+1 honest players. Since honest players will only $P$ valid blocks, condition (1) of safety is true.
	
	Condition (2) of safety can be directly proved by Lemma~\ref{appendix_no_c}. $\hfill \square$

\section{Liveness}
\label{liveness}
Since our potential leaders are elected secretly, the adversaries can not prevent honest players from becoming potential leaders of the least leader score, and can not prevent the block proposed by such an honest player from propagation since gossip peers are also secretly randomly chosen.

We base our proof on a partially synchronous network assumption. And when the network is synchronous, a transaction will fail to reach all honest players in a finite time with negligible probability. If there are always blocks packed by honest players committed, all transactions will be confirmed eventually.

%
%
%
%

\begin{Lem}
\label{appendix_liveness_c}
For any round $r$, there exists a round $r' > r$ and an honest player $p_j$ such that $p_j$ commits some block at height $h = h_{root}+1$ in round $r'$.
\end{Lem}

\para{Proof. }
Without losing generality, we assume that at round $r$, the last committed $B_{root}$ is the same for all honest players. 

(Case 1) No player has $TC$ed. Because the network will become synchronous infinitely, there exists a round where an honest player becomes the potential leader with the least leader score and the network is synchronous. In this round, the proposal round of the block proposed by this honest leader is equal to the local freshness $F$ of all honest players, so it will be prepared and committed by all honest players.

(Case 2) Some player has $TC$ed. Because at most one block can be $TC$ed in a round, if two players $TC$ed different blocks, these two players will have different local $F$. Thus, all players with the largest $F$ will always have the same $B_p$, meaning if any of them becomes the leader in a synchronous round, this $B_p$ will be prepared by all honest players and committed. 
Similar to Case 1, because the network will become synchronous infinitely, there exists a round where an honest player with the largest $F$ becomes the potential leader with the least leader score and the network is synchronous, so its $B_p$ can be proposed, prepared and committed.
%
%
%

To sum up, the lemma is proved. $\hfill \square$

\begin{The}
	\pname protocol achieves liveness.
\end{The}

\para{Proof. }
	By Lemma~\ref{appendix_liveness_c}, the theorem is easily proved. $\hfill \square$

\end{document}